\documentclass[aps,prl,superscriptaddress,amsmath,twocolumn,showpacs,reprint]{revtex4-1}
\usepackage{hangcaption,hhline,psfrag,rotating,amssymb}
\usepackage[hang,nooneline]{subfigure}
\usepackage{dcolumn}
\usepackage{bm}
\usepackage[pdftex]{color}
\usepackage[sort&compress]{natbib}
\usepackage[colorlinks=true,linkcolor=blue,filecolor=blue,urlcolor=blue,citecolor=blue,pdftex=true,plainpages=false]{hyperref}

\graphicspath{{./Dibujos/}}

\newcommand{\order}{{\sf O}}
\newcommand{\ba}{\begin{eqnarray}}
\newcommand{\ea}{\end{eqnarray}}
\newcommand{\be} {\begin{equation}}
\newcommand{\ee} {\end{equation}}

\newcommand{\cpt}{\raise0.4ex\hbox{$\chi$}PT}
\newcommand{\scpt}{S\raise0.4ex\hbox{$\chi$}PT}
\newcommand{\rscpt}{rS\raise0.4ex\hbox{$\chi$}PT}

\begin{document}


\preprint{FERMILAB-PUB-13-504-T}

\title{Determination of \boldmath $|V_{us}|$ from a lattice-QCD calculation of the 
$K\to\pi\ell\nu$ semileptonic form factor with physical quark masses}

\author{A.~Bazavov}
\affiliation{Physics Department, Brookhaven National Laboratory, Upton, New York, USA}
\affiliation{Department of Physics and Astronomy, University of Iowa, Iowa, USA}

\author{C.~Bernard}
\affiliation{Department of Physics, Washington University, St.~Louis, Missouri, USA}

\author{C.M.~Bouchard}
\affiliation{Department of Physics, The Ohio State University, Columbus, Ohio, USA}

\author{C.~DeTar}
\affiliation{Physics Department, University of Utah, Salt Lake City, Utah, USA}

\author{Daping~Du}
\affiliation{Physics Department, University of Illinois, Urbana, Illinois, USA}

\author{A.X.~El-Khadra}
\affiliation{Physics Department, University of Illinois, Urbana, Illinois, USA}
\affiliation{Fermi National Accelerator Laboratory, Batavia, Illinois, USA}

\author{J.~Foley}
\affiliation{Physics Department, University of Utah, Salt Lake City, Utah, USA}

\author{E.D.~Freeland}
\affiliation{Liberal Arts Department, School of the Art Institute of Chicago, 
Chicago, Illinois, USA}

\author{E.~G\'amiz}
\email{megamiz@ugr.es}
\affiliation{CAFPE and Departamento de F\'{\i}sica Te\'orica y del Cosmos,
Universidad de Granada, Granada, Spain}

\author{Steven~Gottlieb}
\affiliation{Department of Physics, Indiana University, Bloomington, Indiana, USA}

\author{U.M.~Heller}
\affiliation{American Physical Society, Ridge, New York, USA}

\author{Jongjeong Kim}
\affiliation{Department of Physics, University of Arizona, Tucson, Arizona, USA}

\author{A.S.~Kronfeld}
\affiliation{Fermi National Accelerator Laboratory, Batavia, Illinois, USA}

\author{J.~Laiho}
\affiliation{SUPA, School of Physics and Astronomy, University of Glasgow, Glasgow, UK}
\affiliation{Department of Physics, Syracuse University, Syracuse, New York, USA}

\author{L.~Levkova}
\affiliation{Physics Department, University of Utah, Salt Lake City, Utah, USA}

\author{P.B.~Mackenzie}
\affiliation{Fermi National Accelerator Laboratory, Batavia, Illinois, USA}

\author{E.T.~Neil}
\affiliation{Fermi National Accelerator Laboratory, Batavia, Illinois, USA}
\affiliation{Department of Physics, University of Colorado, Boulder, Colorado, USA}
\affiliation{RIKEN-BNL Research Center, Brookhaven National Laboratory, Upton, New York, USA}

\author{M.B.~Oktay}
\affiliation{Physics Department, University of Utah, Salt Lake City, Utah, USA}

\author{Si-Wei Qiu}
\affiliation{Physics Department, University of Utah, Salt Lake City, Utah, USA}

\author{J.N.~Simone}
\affiliation{Fermi National Accelerator Laboratory, Batavia, Illinois, USA}

\author{R.~Sugar}
\affiliation{Department of Physics, University of California, Santa Barbara, California, USA}

\author{D.~Toussaint}
\affiliation{Department of Physics, University of Arizona, Tucson, Arizona, USA}

\author{R.S.~Van~de~Water}
\affiliation{Fermi National Accelerator Laboratory, Batavia, Illinois, USA}

\author{Ran Zhou}
\affiliation{Department of Physics, Indiana University, Bloomington, Indiana, USA}
\affiliation{Fermi National Accelerator Laboratory, Batavia, Illinois, USA}

\collaboration{Fermilab Lattice and MILC Collaborations}
\noaffiliation

\date{\today}

\begin{abstract}
We calculate the kaon semileptonic form factor $f_+(0)$ from lattice QCD, working, for the first time, at the
physical light-quark masses.
We use gauge configurations generated by the MILC collaboration with $N_f=2+1+1$ flavors of sea quarks, which
incorporate the effects of dynamical charm quarks as well as those of up, down, and strange.
We employ data at three lattice spacings to extrapolate to the continuum limit.
Our result, $f_+(0)=0.9704(32)$, where the error is the total statistical plus systematic uncertainty added
in quadrature, is the most precise determination to date. 
Combining our result with the latest experimental measurements of $K$ semileptonic decays, one obtains the
Cabibbo-Kobayashi-Maskawa matrix element $|V_{us}|=0.22290(74)(52)$, where the first error is from
$f_+(0)$ and the second one is from experiment. 
In the first-row test of Cabibbo-Kobayashi-Maskawa unitarity, the error stemming from $|V_{us}|$ is now 
comparable to that from $|V_{ud}|$.
\end{abstract}


\maketitle

\phantom{opera}\clearpage

{\it Introduction}: 
The Cabibbo-Kobayashi-Maskawa~\cite{Cabibbo:1963us} (CKM) matrix underpins all quark flavor-changing
interactions in the standard model of particle physics.
Symmetries reduce the number of physical parameters of this $3\times3$ unitary matrix to four.
They can be taken to be $|V_{us}|$, $|V_{ub}|$, $|V_{cb}|$, and 
$\arg\left(V_{ub}^*\right)$, where subscripts denote the quark flavors interacting with the $W$ boson.
The focus of this Letter is to reduce the theoretical uncertainty in the first of these, in a way that
sharpens the test of CKM unitarity from the first row of the matrix.

The test asks whether, or how precisely,
\begin{equation}
    \Delta_u \equiv |V_{ud}|^2 + |V_{us}|^2 + |V_{ub}|^2 - 1
\end{equation}
vanishes.
The CKM matrix elements are determined from, respectively, superallowed nuclear $\beta\,$ decays, 
kaon decays, and $B$-meson decays to charmless final states.
A failure of the test would be evidence for phenomena beyond the standard model.
As it happens, $\Delta_u$ and analogous tests remain in agreement with the CKM paradigm.
Still, the absence of deviations provides stringent constraints on nonstandard phenomena and their 
energy scale~\cite{Cirigliano10}.

Until now, the error
\ba
    \left(\delta\Delta_u\right)^2 &=& 4|V_{ud}|^2\,\left(\delta|V_{ud}|\right)^2 + 
4|V_{us}|^2\,\left(\delta|V_{us}|\right)^2 \nonumber\\
&+& 4|V_{ub}|^2\,\left(\delta|V_{ub}|\right)^2    \label{eq:DeltaCKM}
\ea
has been dominated by the second term, because $|V_{ud}|=0.97425\pm0.00022$ is so 
precise~\cite{Vud08} (the third term is negligible). One can determine $|V_{us}|$ via the axial-vector 
current, i.e., leptonic kaon decays~\cite{Follana:2007uv,%
Durr:2010hr,Aoki:2010dy,Bazavov:2010hj,Laiho:2011np,Bazavov:2013vwa,Dowdall:2013rya}, or via the vector 
current, i.e., semileptonic decays~\cite{ETMC09,RBC10,Kaneko:2012cta,Bazavov:2012cd,Boyle:2013gsa}.
The current precision is at the level of 0.23--0.4\%~\cite{Bazavov:2013vwa,Dowdall:2013rya} for the former,
but only $\sim0.5\%$~\cite{Bazavov:2012cd,Boyle:2013gsa}, for the latter. According to the standard model, 
both approaches should yield the same result, because the $W$-boson current has the structure $V-A$.

For semileptonic decays, the relation between the experimentally measured $K\to\pi\ell\nu(\gamma)$ inclusive
decay width and the CKM matrix element $|V_{us}|$, up to well known overall factors, is~\cite{Cirigliano11}
\begin{equation}
\label{eq:Kl3def}
    \Gamma_{K_{l3 (\gamma)}} \propto 
    \left|V_{us}\right|^2 \left|f_+^{K^0\pi^-}(0)\right|^2 
    \left(1+\delta_{{\rm EM}}^{Kl} + \delta_{{\rm SU(2)}}^{K\pi}\right).
\end{equation}
The quantities $\delta_{{\rm EM}}^{Kl}$ and $\delta_{{\rm SU(2)}}^{K\pi}$ denote long-distance 
electromagnetic and strong isospin-breaking corrections, respectively~\cite{Cirigliano11}. 
The latter is defined as a correction relative to the $K^0$ mode.
The quantity needed from lattice QCD is the vector form factor $f_+(0)$, defined by 
\ba\label{eq:formfac}
\langle \pi (p_\pi)|V^\mu |K(p_K)\rangle & = & 
f_+^{K\pi}(q^2) \left[p_K^\mu + p_\pi^\mu - \frac{m_K^2-m_\pi^2}{q^2}
q^\mu\right]\nonumber\\&& { } + f_0^{K\pi}(q^2)\frac{m_K^2-m_\pi^2}{q^2}q^\mu ,
\ea
where $V^\mu=\bar s\gamma^\mu u$ and $q=p_K-p_\pi$ is the momentum transfer.

We previously~\cite{Bazavov:2012cd} presented a lattice-QCD calculation of $f_+(0)$ using the $N_f=2+1$ 
gauge-field configurations generated by the MILC Collaboration.
The RBC/UKQCD Collaboration presented 
an independent calculation~\cite{Boyle:2013gsa}, using a different set of $N_f=2+1$ gauge-field configurations. 
Even though both works reduce the error on $|V_{us}|$ from $f_+(0)$ to $\sim 0.5$\%, it is still
roughly two times larger than the experimental uncertainty from $\Gamma_{K_{l3(\gamma)}}$.

Before, our dominant systematic uncertainty came from the chiral extrapolation of light-quark masses from
their simulation values to the physical point~\cite{Bazavov:2012cd}.
Here, we reduce this uncertainty by a factor of five with data directly at the physical light-quark mass.
Thus, the extrapolation becomes an interpolation.
We work with a subset of the $N_f=2+1+1$ ensembles generated (again) by the MILC
Collaboration~\cite{HISQensembles}.
The new ensembles use an action for the sea quarks with three-times smaller discretization effects.
We now use three different lattice spacings, instead of only two. 
In these ensembles, the strange sea-quark masses are much better tuned than before, reducing another important
uncertainty in Ref.~\cite{Bazavov:2012cd}.
Finally, the new ensembles include the effects of charm quarks in the sea.

{\it Simulation details and statistical errors}: %
We largely follow the strategy of Ref.~\cite{Bazavov:2012cd}.
Hence, this Letter only summarizes the main features and points out the differences.
We refer the reader to Ref.~\cite{Bazavov:2012cd} for details of our methodology and to 
Ref.~\cite{Lattice2013} for technical details of the current numerical work.

We obtain the form factor using the relation~\cite{HPQCD_Dtopi} 
\begin{equation}
\label{eq_WI}
    f_+(0) = f_0(0) = \frac{m_s-m_l}{m_K^2-m_\pi^2}\langle \pi(p_\pi)|s\bar u |K(p_K)\rangle\,.
\end{equation}
The last expression requires no renormalization and allows us to extract the form factor from three-point 
correlation functions with less noise than Eq.~(\ref{eq:formfac}). The momentum of the pion, $p_\pi$, 
is adjusted via partially twisted boundary conditions~\cite{Bedaque:2004ax,tbc}, such that $q^2=0$.

Table~\ref{tab:ensemblesHisq} shows the simulation parameters of the ensembles used here~\cite{HISQensembles}.
\begin{table}[tp]
    \centering
\caption{Parameters of the $N_f=2+1+1$ gauge-field ensembles and correlation functions generated in this work.
$N_\text{conf}$ is the number of configurations included, $N_\text{src}$ the number of time 
sources used on each configuration, and $L$ the spatial size of the lattice. Pion masses (fourth and fifth 
columns) are given in MeV. Further information, including the light and charm quark masses, can be found in
Ref.~\cite{HISQensembles}.}
\label{tab:ensemblesHisq}
\begin{tabular}{cllcccrc}
\hline\hline
      $\approx a$ (fm) & ~$am_s^\text{sea}$ & ~$am_s^\text{val}$ & 
$m_{\pi}^P$ &  $m_{\pi}^\text{RMS}$ & $m_\pi L$ 
& $N_\text{conf}\!$ & $N_\text{src}$ \\
\hline
0.15 & 0.0647 & 0.06905 & 133 & 311 & 3.30 & $1000$ & $4$   \\
\hline
0.12 & 0.0509 & 0.0535  & 309 & 370 & 4.54 & $1053$ & $8$ \\
     & 0.0507 & 0.053   & 215 & 294 & 4.29 &  $993$ & $4$ \\
     & 0.0507 & 0.053   & 215 & 294 & 5.36 &  $391$ & $4$  \\
     & 0.0507 & 0.0531  & 133 & 241 & 3.88 &  $945$ & $8$  \\
\hline
0.09 & 0.037  & 0.038   & 312 & 332 & 4.50 &  $775$ & $4$ \\
     & 0.0363 & 0.038   & 215 & 244 & 4.71 &  $853$ & $4$ \\
     & 0.0363 & 0.0363  & 128 & 173 & 3.66 &  $621$ & $4$ \\
\hline
0.06 & 0.024  & 0.024   & 319 & 323 & 4.51 &  $362$ & $4$ \\
\hline\hline
\end{tabular}
\end{table}
These ensembles use a one-loop Symanzik-improved gauge action for the gluons~\cite{gauge1,gauge2}, and the
highly-improved staggered-quark (HISQ) action~\cite{Hisqaction} for the $u$, $d$, $s$, and $c$ quarks in the
sea. The HISQ sea quarks were simulated with the fourth-root procedure for eliminating extra quark species
(often called tastes) arising from fermion doubling~\cite{Marinari:1981qf,Durr:2003xs,Follana:2004sz,%
DHW04,Shamir04,Bernard06,Bernard:2006vv,BGS08,Adams:2008db,Donald:2011if}.

We study data at four different values of the lattice spacing. The number of configurations analyzed at 
$a\approx0.06~{\rm fm}$ is too small to remove autocorrelation effects in a controlled way, so this data set 
is not used in the central fit but as a cross-check of discretization effects.

The strange and charmed masses are always near their physical values. In most cases, however, a better 
tuning of $m_s$ became available before computing the matrix element in Eq.~(\ref{eq_WI}).
We have chosen the better tuned value for the valence quarks, hence the different values of~$m_s^\text{val}$
and~$m_s^\text{sea}$. The up and down sea-quark masses are taken to be the same: $m_l=0.2m_s$, $0.1m_s$, 
or $m_s/27$. The last corresponds very nearly to the physical pion mass, $135~{\rm MeV}$.  
We include data at heavier-than-physical pion masses to further control the chiral-continuum fit.

While the column labeled by $m_\pi^P$ in Table~\ref{tab:ensemblesHisq} corresponds to the 
valence pion, the root-mean-squared pion mass, $m_\pi^\text{RMS}$, provides a measure of the dominant 
discretization effects, due to lattice-artifact interactions between staggered quarks of different tastes.
These taste splittings, are of order $\alpha_s^2a^2$ for the HISQ action, where $\alpha_s$ is the strong
coupling at a scale around $\pi/a$. They decrease rapidly with the lattice spacing, as can be seen from 
the difference of the fifth and fourth columns in Table~\ref{tab:ensemblesHisq}.

We obtain both hadronic matrix elements and meson energies from combined fits of two-point and
three-point correlation functions. The structure of these three-point functions is the 
same as in Ref.~\cite{Bazavov:2012cd}, but here we only include moving $\pi$ data~\cite{Lattice2013}.
The correlation function fits include ground states, and excited and opposite-parity
states~\cite{Bazavov:2012cd}.

Our correlation-function fits are stable under variations of the number of states, time ranges, source-sink
separations, and other aspects of the fits.
The central values and statistical errors are shown as a function of the light quark mass in
Fig.~\ref{fig:ChPTcentral}, which is discussed in more detail below.
Numerical values are given in Ref.~\cite{Lattice2013}.
Within the statistical errors of relative size $\sim0.2$--0.4\%, the data show no discretization effects
except possibly at $0.15$~fm. 

{\it Chiral interpolation and continuum extrapolation}: %
\begin{figure}[bp]
    \centering
    \vspace*{-0.7cm}
    \includegraphics[width=0.5\textwidth]{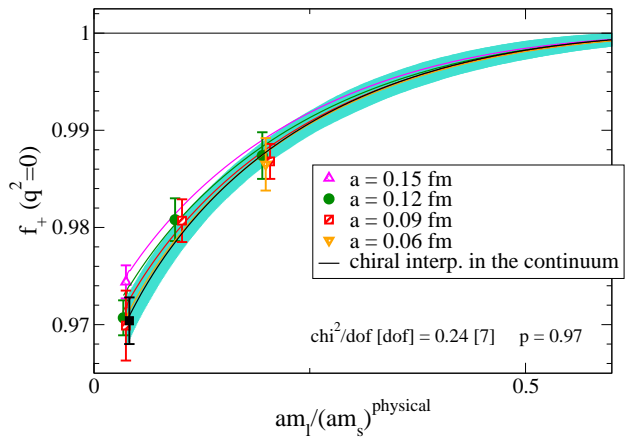}
    \caption{Form factor $f_+(0)$ vs.~light-quark mass.
        Errors shown are statistical only, obtained from 500 bootstraps.
        Different symbols and colors denote different lattice spacings, and the corresponding colored lines 
        show the chiral interpolation at fixed lattice spacing.
        The solid black line is the interpolation in the light-quark mass, keeping $m_s$ equal to its 
        physical value, and turning off all discretization effects.
        The turquoise error band includes statistical, discretization, and higher order chiral errors, as 
        explained in the text.}
    \label{fig:ChPTcentral}
\end{figure}
Even though we have data at the physical quark masses, we include data at larger $m_\pi$.
The use of chiral perturbation theory ($\chi$PT) and data at different masses allows us to correct for
small mistunings of the light- and strange-quark masses, as well as for partially quenched effects due to
$m_s^\text{val}\ne m_s^\text{sea}$. 
In addition, these data are very precise and help to reduce the final statistical error. 
Furthermore, the dominant discretization effects, are 
well-described by the $\chi$PT formula~\cite{Lee:1999zxa,SCHPT}, so they are removed when taking the
continuum limit.

In $\chi$PT, the form factor $f_+(0)$ can be written as $f_+(0) = 1 + f_2 + f_4 + \ldots$.
In continuum QCD, the Ademollo-Gatto theorem~\cite{AGtheorem} ensures that the $\order(p^{2i})$ chiral
corrections $f_{2i}$ tend to zero in the SU(3) limit as $(m_K^2-m_\pi^2)^2$.
In particular, $f_2$ is completely fixed in terms of well-known quantities.
At finite lattice spacing, however, violations of the Ademollo-Gatto theorem arise from discretization
effects in the dispersion relation needed to derive the relation in Eq.~(\ref{eq_WI}).

We perform the interpolation to the physical masses and the continuum using next-to-next-to-leading-order
(NNLO) continuum $\chi$PT~\cite{BT03}, supplemented by next-to-leading-order (NLO) partially quenched,
staggered $\chi$PT~\cite{KtopilnuSChPT}.
Because we observe almost no lattice-spacing dependence in our data, discretization effects in higher-loop
$\chi$PT should be negligible.
After removing the dominant discretization effects with S$\chi$PT, the remaining ones, which stem from
violations of the continuum dispersion relation and higher orders taste-splitting effects, are of order
$\alpha_sa^2$, $a^4$, $(m_K^2-m_\pi^2)^2\alpha_s a^2$, and $(m_K^2-m_\pi^2)^2\alpha_s^2a^2$.
We introduce fit parameters for these terms---$K_1$, $K_3$, $K_2$, and $K'_2$, respectively---and
take the functional form
\ba
\label{eq:ChPTtwoloop} 
    f_+(0) & = & 1 + f_2^{{\rm PQS\chi PT}}(a) + K_1\,a^2\sqrt{\bar{\Delta}} + K_3\,a^4
            + f_4^\text{cont} \nonumber \\
        & + & (m_\pi^2-m_K^2)^2 \left[C_6+K_2a^2\sqrt{\bar{\Delta}}
            + K_2'a^2\bar\Delta\right. \nonumber \\
        & & \hphantom{(m_\pi^2-m_K^2)^2N} \left. \vphantom{\sqrt{\bar\Delta}} 
            + C_8\, m_\pi^2 + C_{10}\,m_\pi^4\right],
\ea
where $f_2^{{\rm PQS\chi PT}}(a)$ is the NLO partially quenched staggered $\chi$PT expression including leading
isospin corrections~\cite{Gasser:1984ux}, $f_4^\text{cont}$ is the sum of the NNLO continuum chiral logarithms, 
and $a^2\bar\Delta$ is the average taste splitting, with $\bar\Delta$ used as a proxy for $\alpha_s^2$. 
The analytic term $C_6$ is related to a combination of low-energy constants in continuum $\chi$PT, and $C_8$
and $C_{10}$ are fit parameters that parametrize chiral corrections at N${}^3$LO and N${}^4$LO, respectively.
We take the taste splittings from Ref.~\cite{HISQensembles} and set the rest of the inputs in the same way as
in Ref.~\cite{Bazavov:2012cd}. The fit parameters are constrained with Bayesian techniques. 
We fix their prior widths using power counting arguments, except for $K'_2$, where we triple
the power-counting width, since it is the numerically dominant term in our fits~\cite{Lattice2013}.
Using Eq.~(\ref{eq:ChPTtwoloop}) expressed in terms of meson masses, we interpolate to physical pion and kaon
masses with electromagnetic effects removed~\cite{FLAG2,Basak:2013iw}: $m_{\pi^+}^{\rm QCD}=135.0$~MeV, $m_{K^0}^{\rm QCD}\approx m_{K^0}^{\rm phys}=497.7$~MeV, and $m_{K^+}^{\rm QCD}=491.6$~MeV.
The last value enters only in $f_2$.

We estimate the statistical errors by generating a set of 500 pseudoensembles via 
the bootstrap method, and repeating the fit on each pseudoensemble. The result from the
chiral and continuum interpolation/extrapolation is $f_+(0)=0.9704(24)$, which is shown in
Fig.~\ref{fig:ChPTcentral}.
The fits cannot precisely determine the coefficients $K_i$ in Eq.~(\ref{eq:ChPTtwoloop}), 
since only the $a\approx 0.15~{\rm fm}$ point appears to show any discretization effects. 
We examine this issue via fits with fewer parameters, including 
one-by-one the analytical $a^2$ terms in Eq.~(\ref{eq:ChPTtwoloop}), and excluding higher order
chiral terms (third line in Eq.~(\ref{eq:ChPTtwoloop})) to make the comparison cleaner. The 
results of these fits are shown in Table~\ref{tab:asqstability}. 
We find no difference except when all of the discretization effects are omitted. 
Something similar happens with the addition of higher order chiral terms to the fit function. 
Adding a N${}^3$LO term slightly changes the central value and increases the error from 
0.9703(23) to 0.9704(24). Adding a N${}^4$LO term does not change either 
the central value or the error. The alternate fits with additional discretization terms 
and/or chiral terms show that fit errors are saturated.  We thus consider the error from 
the chiral and continuum interpolation/extrapolation with the fit function in 
Eq.~(\ref{eq:ChPTtwoloop}), $f_+(0)=0.9704(24)$, as 
the total statistical+discretization+chiral interpolation error. 
The increase in the error when adding a N${}^3$LO term,
 $~0.0004$, gives a measure of the chiral interpolation error, five times
smaller than in our previous work~\cite{Bazavov:2012cd}, thanks to the inclusion 
of data at physical quark masses. We discuss further tests of 
the robustness of this Bayesian error estimate strategy in Ref.~\cite{Lattice2013}.

\begin{table}[tp]
    \caption{Stability of the continuum extrapolation with omission of discretization terms, in the notation 
        of Eq.~(\ref{eq:ChPTtwoloop}).}
    \label{tab:asqstability}
    \begin{tabular}{lccc}
    \hline\hline
    \multicolumn{1}{c}{Parameters omitted}       &  $f_+(0)$  & $\chi^2/\text{dof}$ & $p$  \\
    \hline
    $C_8$, $C_{10}$, $K_1$, $K_2$, $K_3$, $K'_2$ & 0.9714(12) &         0.27        & 0.97 \\
    $C_8$, $C_{10}$, $K_1$, $K_2$, $K_3$         & 0.9703(23) &         0.24        & 0.97 \\
    $C_8$, $C_{10}$,        $K_2$, $K_3$         & 0.9703(23) &         0.24        & 0.97 \\
    $C_8$, $C_{10}$,               $K_3$         & 0.9703(23) &         0.24        & 0.97 \\
    $C_8$, $C_{10}$                              & 0.9703(23) &         0.24        & 0.97 \\
    \hline
    Central fit: full Eq.~(\ref{eq:ChPTtwoloop}) & 0.9704(24) & 0.24 & 0.97 \\
    \hline\hline
    \end{tabular}
\end{table}

Although we omit it from the chiral and continuum interpolation/extrapolation, 
we also show data on an ensemble with a smaller lattice spacing, $a\approx 0.06~{\rm fm}$, and 
$m_l^\text{sea}=0.2m_s^\text{sea}$, the (orange) down-pointing triangle in Fig.~\ref{fig:ChPTcentral}.
It lies on top of the results for the other lattice spacings, confirming that discretization effects 
are much smaller than statistical errors. The same conclusion follows from the fact that the red line 
in Fig.~\ref{fig:ChPTcentral}, for $a\approx 0.09~{\rm fm}$, is very close to the continuum one. The remaining 
significant sources of systematic uncertainty are given in Table~\ref{tab:errorbudget}. We estimate the error 
due to including partially quenched effects only at one loop by the shift in the 
final result when using $m_s^{\text{val}}$ or $m_s^{\text{sea}}$ in the NNLO chiral 
logarithmic function, $f_4^{{\rm cont}}$. To convert dimensionful quantities from lattice 
to physical units, we use the scale $r_1=0.3117(22)~{\rm fm}$~\cite{decayconstants2011} 
obtained from the static-quark potential~\cite{r11,r12}. The form factor, being a 
dimensionless quantity, depends on the scale only via the input parameters. 
Propagating the uncertainty in the scale 
through to $f_+(0)$ yields the entry shown in Table~\ref{tab:errorbudget}. 
For an estimate of the finite volume error we compare our data obtained with two 
different spatial volumes and other parameters at $a\approx 0.12$~fm fixed. The difference 
is about half of the statistical error, so we take the finite 
volume error to be the full size of the statistical error. Finally, 
we estimate the error from the NNLO and higher order isospin corrections to the 
$K^0\pi^+$ mode by taking twice the difference between the NNLO contribution 
to $f_+(0)$ with and without isospin corrections~\cite{Bijnens:2007xa}.  See Ref.~\cite{Lattice2013} 
for more details.

{\it Final result and conclusions}: Our final result for the vector form factor is
\ba\label{eq:fplus}
f_+(0) = 0.9704(24)(22) = 0.9704(32), 
\ea
where the first error is from the chiral-continuum fit, and the second the sum in 
quadrature of the other systematic errors listed in Table~\ref{tab:errorbudget}.  
This result is the most precise calculation of $f_+(0)$ to date 
and the first to include data at physical light-quark masses. 
It agrees with the previous results of Refs.~\cite{Bazavov:2012cd,Boyle:2013gsa}, 
with a reduced total uncertainty of 0.33\%. 

Using the latest average of experimental results for $K$ semileptonic decays,
$|V_{us}|f_+(0)=0.2163(5)$~\cite{MoulsonCKM12}, and the form factor in Eq.~(\ref{eq:fplus}), one obtains
\begin{equation}
    |V_{us}| = 0.22290(74)_{f_+(0)}(52)_\text{expt}=0.22290(90).
    \label{eq:Vus}
\end{equation}
The unitarity test becomes
\begin{equation}
    \Delta_u=-0.00115(40)_{V_{us}}(43)_{V_{ud}},
\end{equation}
i.e., the error on $\Delta_u$ from $|V_{us}|$ is now slightly smaller than that from $|V_{ud}|$.
Combining the two errors, one sees a $\sim2\sigma$ tension with unitarity. 
Recall that the semileptonic decay proceeds through the vector current;
the uncertainty of $|V_{us}|/|V_{ud}|$ from the axial-vector current, via leptonic pion and kaon decays and 
the ratio $f_K/f_\pi$ \cite{Dowdall:2013rya} already results in a value of $\Delta_u$ with smaller error. 
As emphasized above, it is important to carry out the test with both currents.

\begin{table}[tp]
    \centering
\caption{\label{tab:errorbudget} Error budget for $f_+(0)$ in percent.} 
\begin{tabular}{lc}
\hline
 \hline
Source of uncertainty &  Error $f_+(0)$ (\%) \\
\hline
Stat. + disc. + chiral inter. & $0.24$ \\
$m_s^\text{val}\ne m_s^\text{sea}$ & $0.03$ \\
Scale $r_1$ & $0.08$ \\
Finite volume & $0.2$  \\
Isospin & $0.016$ \\ 
\hline
Total Error & $0.33$ \\
\hline\hline
\end{tabular}
\end{table}

In summary, with the HISQ $N_f=2+1+1$ ensembles, we have reduced the uncertainties on $|V_{us}|$ from the
chiral interpolation and discretization effects.
The main remaining sources of error are Monte Carlo statistics and finite-volume effects.
In order to reach the final target of 0.2\% precision required by experiment, we are increasing statistics
and deriving the finite-volume corrections at one-loop in partially quenched staggered $\chi$PT with twisted
boundary conditions~\cite{ChPT_FV}.

{\it Acknowledgements}: We thank Christine Davies for useful discussions. 
We thank Johan Bijnens for making his NLO partially quenched $\chi$PT and NNLO
full QCD $\chi$PT codes available to us. A.X.K. thanks the Fermilab theory group for 
hospitality while this work was finalized. 
Computations for this work were carried out with resources provided by 
the USQCD Collaboration, the Argonne Leadership Computing Facility and
the National Energy Research Scientific Computing Center, which
are funded by the Office of Science of the United States Department of Energy; and with resources
provided by
the National Center for Atmospheric Research,
the National Center for Supercomputing Applications,
the National Institute for Computational Science,
and the Texas Advanced Computing Center,
which are funded through the National Science Foundation's Teragrid/XSEDE and Blue Waters Programs. 
This work was supported in part by the U.S. Department of Energy under Grants
No.~DE-FG02-91ER40628 (C.B.),
No.~DE-FC02-06ER41446 (C.D., J.F., L.L., M.B.O.),
No.~DE-FG02-91ER40661 (S.G., R.Z.),
No.~DOE DE-FG02-13ER42001 (D.D., A.X.K.),
No.~DE-FG02-04ER-41298 (J.K., D.T.);
by the National Science Foundation under Grants
No.~PHY-1067881, No.~PHY-0757333, No.~PHY-0703296 (C.D., J.F., L.L., M.B.O.),
No.~PHY-0757035 (R.S.); 
by the URA Visiting Scholars' program (A.X.K.);
by the Science and Technology Facilities Council and the Scottish Universities
Physics Alliance (J.L.);
by the MINECO (Spain) under Grants FPA2010-16696, FPA2006-05294, and \emph{Ram\'on y Cajal} program (E.G.); 
by Junta de Andaluc\'{\i}a (Spain) under Grants FQM-101 and FQM-6552 (E.G.); 
and by European Commission under Grant No. PCIG10-GA-2011-303781 (E.G.). 
This manuscript has been co-authored by an employee of Brookhaven Science Associates,
LLC, under Contract No.~DE-AC02-98CH10886 with the U.S. Department of Energy.
Fermilab is operated by Fermi Research Alliance, LLC, under Contract
No.~DE-AC02-07CH11359 with the U.S. Department of Energy.


\end{document}